\documentclass[final,5p,times,twocolumn]{elsarticle}

\usepackage{graphicx}
\usepackage{amsmath}
\usepackage{amsfonts}
\usepackage{amssymb}

\usepackage{lineno}

\journal{Journal of Theoretical Biology}

\begin{document}

\begin{frontmatter}


\title{Seven rules to avoid the tragedy of the commons}



\author{Yohsuke Murase}
\address{RIKEN Center for Computational Science, Kobe, Hyogo
650-0047, Japan}
\ead{yohsuke.murase@gmail.com}

\author{Seung Ki Baek\corref{cor1}}
\address{Department of Physics, Pukyong National University, Busan 48513, Korea}
\ead{seungki@pknu.ac.kr}

\begin{abstract}
Cooperation among self-interested players in a social dilemma is fragile and easily interrupted by mistakes. In this work, we study the repeated $n$-person public-goods game and search for a strategy that forms a cooperative Nash equilibrium in the presence of implementation error with a guarantee that the resulting payoff will be no less than any of the co-players'. By enumerating strategic possibilities for $n=3$, we show that such a strategy indeed exists when its memory length $m$ equals three. It means that a deterministic strategy can be publicly employed to stabilize cooperation against error with avoiding the risk of being exploited. We furthermore show that, for general $n$-person public-goods game, $m \geq n$ is necessary to satisfy the above criteria.
\end{abstract}

\begin{keyword}
Evolution of cooperation \sep Public-goods game \sep Reciprocity


\end{keyword}

\end{frontmatter}



\section*{Introduction}

Conflicts between individual and collective interests are observed across a
variety of fields from genetics to international politics. For example, genes
can inflict damage to other genes in the same genome for spreading at a higher
rate even if it threatens the
`host'~\cite{haig1993genetic,hurst1996genetic,burt2009genes,archetti2003selfish,friberg2008cut}.
Some microorganisms spend energy to produce chemicals that are beneficial for
the whole population, which can be exploited by non-cooperating
mutants~\cite{crespi2001evolution,velicer2003social,greig2004prisoner,rankin2007tragedy,gore2009snowdrift}.
The same conflict exists in cooperative mammals when they hunt in group or stand
guard against
predators~\cite{bednarz1988cooperative,packer1990lions,clutton1999selfish}.
One of such cooperative species is {\it Homo sapiens}, the political
animal~\cite{kollock1998social}. Not surprisingly, human societies have
constantly experienced the \emph{Tragedy of the
Commons}~\cite{hardin1968tragedy,ostrom1999revisiting,baer2000equity,dietz2003struggle}
and struggled to preserve common goods against it. There are a few
conventional ways to achieve this goal: The first is the `Leviathan' solution
involving governmental
regulations~\cite{ophuls1973leviathan,heilbroner1980inquiry,carruthers1981economic}.
The next is the market mechanism dealing with the common resource as private
property~\cite{demsetz1974toward,smith1981resolving,sinn1984common}. The third
one is institutional design for collective actions of civil
society~\cite{ostrom1990governing} such as
punishment~\cite{boyd1992punishment,fehr2002altruistic,boyd2003evolution,de2004neural,gurerk2006competitive,hauert2007via,szolnoki2015antisocial,chen2015competition,szolnoki2017second}
combined with a reputation
system~\cite{brandt2003punishment,rockenbach2006efficient}.
However, the above answers are becoming hard to justify on a global scale
because there is no world government~\cite{paavola2012climate}, market failure
is likely to occur~\cite{stern2007economics}, and institutionalism
tends to fail for a large group of people~\cite{araral2014ostrom}.
The question is then what can be done about players that refuse to contribute.

To give a concrete form to this question, let us consider the $n$-person
public-goods (PG)
game~\cite{boyd1988evolution,milinski2006stabilizing,milinski2008collective,perc2017statistical}.
In our setting, each player may either cooperate ($c$) by contributing a private
token to a public pool or defect ($d$) by refusing it.
The tokens are then multiplied by a factor $\rho$ and equally distributed to the
players. The multiplication factor $\rho$ must be greater than one and smaller
than the number of players to describe the conflict between individual and
collective interests. A player's payoff is then defined as
\begin{equation}
  \begin{cases}
    \frac{\rho n_c}{n} & \text{when the player cooperates ($c$)} \\
    1 + \frac{\rho n_c}{n} & \text{when the player defects ($d$)} \\
  \end{cases},
\end{equation}
where $n_c$ is the number of cooperators including the focal player.
Everyone prefers $d$, and zero contribution is a Nash equilibrium
of this one-shot game. In many circumstances, however, the players are bound to
interact repeatedly for a long time. Under the shadow of the future, it becomes
possible to devise strategies conditioned on previous interactions, whereby
reciprocity~\cite{trivers1971evolution,hamilton1981evolution,frank2018detecting}
comes into play in organizing collective efforts for the public pool. For
example, a generalized version of tit-for-tat (TFT) forms a Nash equilibrium if
error probability is strictly
zero~\cite{boyd1988evolution,lichbach1992repeated} in a noiseless
environment~\cite{fudenberg1998theory,szabo2007evolutionary,szolnoki2009topology}.

\begin{figure}
\begin{center}
  \includegraphics[width=0.8\columnwidth]{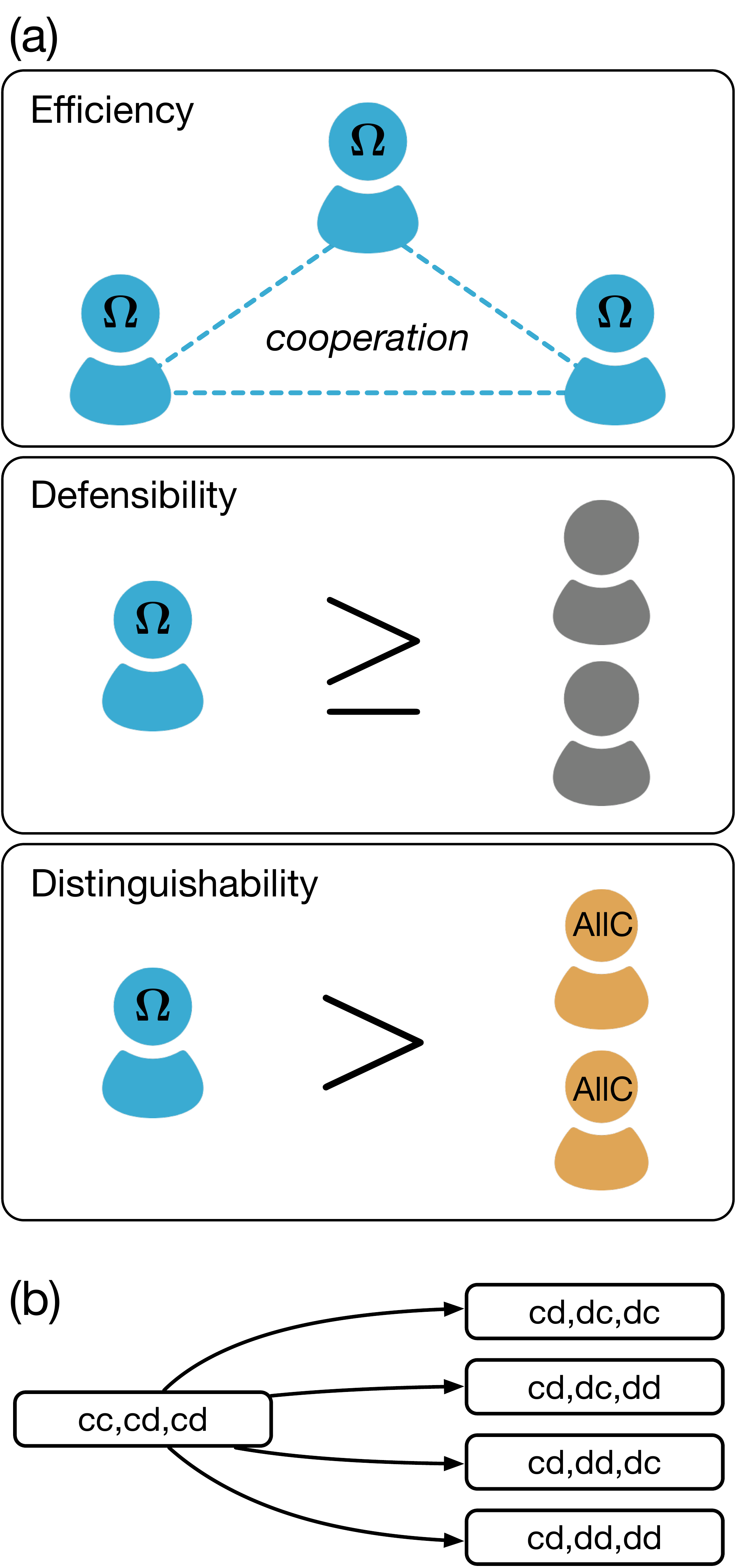}
  \caption{\label{fig:three_conditions_sample_graph}
  (a) Illustration of the three criteria imposed on a strategy $\Omega$.
  [i] Efficiency: Mutual cooperation is realized when all players use $\Omega$.
  [ii] Defensibility: It is guaranteed that a player using $\Omega$ never
  has a lower long-term payoff than those of the co-players whatever strategies
  they use.
  [iii] Distinguishability: When the co-players are naive cooperators,
  a player using $\Omega$ has a strictly higher payoff than the others'.
  (b) An example of a transition graph.
  Suppose that Alice takes $d$ at the state of $(cc,cd,cd)$.
  There are four next possible states $(cd,dc,dc)$, $(cd,dc,dd)$, $(cd,dd,dc)$,
  and $(cd,dd,dd)$ depending on the moves of Bob and Charlie.
  In this manner, a memory-2 strategy is represented by a graph having $2^6$
  nodes, each of which has $4$ outgoing links.
  }
  \end{center}
\end{figure}

In terms of the repeated PG game, our question is the following: What can we
advise players of this game in the presence of implementation error, if they
wish to achieve full cooperation without being exploited repeatedly by others?
The direct $n$-person generalization of TFT cannot be an answer because it
results in a series of wasteful retaliation when someone defects by mistake.
An all-or-none strategy~\cite{pinheiro2014evolution}, a generalization of
win-stay-lose-shift~\cite{kraines1989pavlov,nowak1993strategy,posch1999win,liu2012win}, is a strong
candidate for our purpose because it constitutes a
subgame perfect Nash equilibrium if the benefit-cost ratio of cooperation is
sufficiently high~\cite{hilbe2017memory}. Unfortunately, the cost-benefit
analysis {\it per se} is often a difficult issue in practice. Another
drawback is that this strategy systematically yields a higher payoff to its
co-player who plays unconditional defection (AllD).
Although the idea of the zero-determinant (ZD)
strategies~\cite{press2012iterated,hilbe2014cooperation,szolnoki2014defection,szolnoki2014evolution} may help to have
control over the situation in the $n$-person case~\cite{pan2015zero},
the probabilistic retaliation prescribed by most ZD strategies can
hardly be an available option when it comes to
policymaking~\cite{dror1983public}. Indirect
reciprocity~\cite{sugden1986economics,nowak1998evolution,ohtsuki2004should,ohtsuki2009indirect,panchanathan2004indirect,suzuki2005reputation,suzuki2007evolution,suzuki2007three,mcnamara2015reputation,nax2015stability}
is also difficult to bring into action because it requires an agreement on every
player's reputation within a time scale of successive interactions.
To sum up, we need a deterministic strategy of direct reciprocity such that
solves the repeated PG game with an arbitrary multiplication factor $\rho \in
(1,n)$ and nonzero error probability $e >0$. It is reasonable
to say that the players may safely adopt a certain strategy $\Omega$ if it
satisfies the following three criteria~\cite{yi2017combination} (see
Fig.~\ref{fig:three_conditions_sample_graph}(a)).
\begin{enumerate}
  \item
    Efficiency: Mutual cooperation is achieved with probability one as
    error probability $e$ approaches zero, when all players use this strategy
    $\Omega$.
    Mathematically speaking, we consider a strategy profile $\mathcal{P}
    =\{s_1, s_2, \ldots, s_i, \ldots, s_n\}$ of $n$ players, and the relevant
    observable is player $i$'s long-term payoff, defined as
    \begin{equation}
    f_{i} \equiv \lim_{T\to\infty} \frac{1}{T} \sum_{t=0}^{T-1} F_{i}^{(t)},
    \end{equation}
    where $F_{i}^{(t)}$ denotes the instantaneous payoff at time $t$ that player
    $i$ receives with $s_i$. When everyone uses $\Omega$ with $e>0$,
    the Markovian dynamics of strategic interaction converges to a
    unique stationary distribution, from which $f_i$ is readily
    calculated for a given strategy
    profile~\cite{nowak1990stochastic,nowak1995automata}.
    The efficiency criterion essentially means that $\lim_{e\to 0^+}f_{i} =
    \rho$ when $\mathcal{P} = \mathcal{P}_{\Omega} \equiv
    \{ \Omega, \Omega,\ldots, \Omega \}$.
  \item
    Defensibility: The strategy $\Omega$ ensures that none of the co-players can
    obtain higher long-term payoffs against $\Omega$ regardless of the
    co-players' strategies and initial state when $e=0$. This condition assures
    $\lim_{e \to 0^+} \left( f_i - f_j \right) \ge 0$, where $s_i = \Omega$ and
    $j$ denotes any possible co-player of $i$.
    When combined with efficiency, this criterion is strong enough
    for $\mathcal{P}_{\Omega}$ to be a cooperative Nash equilibrium, in which
    $\lim_{e \to 0^+} f_i = \rho$ for every player $i$. To verify this
    statement,
    suppose that a player, say $j$, unilaterally switches to another strategy
    while the others keep using $\Omega$. Player $j$'s resulting long-term
    payoff is denoted as $f_j'$, and each of the other $\Omega$-using players
    obtains a certain payoff $\phi$, which may not equal $\rho$. According to
    the Pareto optimality of $S_{\Omega}$, the total payoff of the $n$ players
    becomes less than or equal to that of $\mathcal{P}_{\Omega}$, i.e., $(n-1)
    \phi + f_j' \le n \rho$. Defensibility means that $\phi \geq f_{j}'$ for
    $j$'s every possible choice, which leads to the conclusion that $f_j' \le
    \rho$.
  \item
    Distinguishability: If $s_i = \Omega$ and all its co-players are
    unconditional cooperators (AllC), player $i$ can exploit them to earn a
    strictly higher long-term payoff than the co-players'. That is,
    $f_{i} > f_{j}$ when $j$ is an AllC player.
    This may be sharpened further by requiring the same inequality for
    every possible number of $\Omega$-players between one and $n-1$,
    but such refinement turns out to be irrelevant in this work.
    This criterion is introduced to suppress invasion of AllC due to neutral
    drift~\cite{imhof2005evolutionary,imhof2007tit,imhof2010stochastic}.
    It should be noted, however, that this criterion does not fully eliminate
    the possibility of a second-order drift via a third strategy between
    $\Omega$ and AllC.
\end{enumerate}
Note the seemingly conflicting requirements expressed in these criteria:
$\Omega$ must recover cooperation from erroneous defection while protecting
itself from malicious ones. It is very doubtful that one can tell
other players' intentions, however, especially when they have longer memories
and better computational power. Worse is that they may even conspire together to
entrap our focal player. The dilemma between efficiency and defensibility is so
severe that one often feels almost forced to compromise one of them,
but we have to ask ourselves whether they are really mutually exclusive.

In fact, it is known that the criteria can be met in the Prisoner's
Dilemma (PD) game, an equivalent of the two-person PG game. The resulting
strategy is based on TFT but able to correct the player's own
error~\cite{yi2017combination}. It is tempting to apply this strategy to
the $n$-person game,
e.g., by reducing the situation to an effective two-person game between one
and the other players. However, this idea does not work when $n > 2$ for the
following reason: Suppose that everyone has adopted the strategy. When someone
made a mistake, the other players will respond by taking $d$ to defend
themselves. Although the first player tries to redeem the mistake, the point is
that the other $n-1$ players see each other choosing $d$. As long as some other
players are defecting, the strategy will advise against returning to $c$ for the
sake of defensibility, so it fails to reestablish cooperation.
This `observer' effect illustrates a fundamental difficulty of the $n$-person
game when $n > 2$.

Now, our question on the Tragedy of the Commons boils down to whether it is
possible to meet the three criteria of efficiency, defensibility, and
distinguishability for $n>2$.
In this work, we report two findings: First, when the number
of players is $n=3$, we can explicitly construct $\Omega$ whose memory length
$m$ is three. Second, we show for the general $n$-person case that
$m$ must be greater than or equal to $n$ for a strategy to satisfy the
efficiency and the defensibility criteria simultaneously.
In other words, the Tragedy of the Commons among three players can be safely
solved for an arbitrary multiplication factor, in the sense of a cooperative
Nash equilibrium, when the error probability is vanishingly small yet nonzero.
At the same time, such a solution will become more and more intricate as the
number of players increases.

\section*{Methods}

Let us explain how to check the above criteria in the three-person
PG game by means of direct enumeration. First of all, we have to define
the game. We will denote the three players as Alice, Bob, and Charlie,
respectively. The payoff matrix from Alice's point of view is defined as
\begin{equation}
M \equiv \left(
  \begin{array}{c|ccc}
    \ & 0 & 1 & 2  \\\hline
    c & \rho & \frac{2}{3}\rho & \frac{1}{3}\rho \\
    d & 1+\frac{2}{3}\rho & 1+\frac{1}{3}\rho & 1
  \end{array}
\right),
\end{equation}
where the column indices represent the number of defectors among Bob and
Charlie.
The next step is to choose an appropriate strategy space. It is common to
classify strategies according to their memory length
$m$~\cite{baek2016comparing}. For example, if a strategy has $m=2$, it refers to
two previous time steps to make a decision at time $t$.
The players' memory state in total can be written as
$S_t = ( A_{t-2} A_{t-1}, B_{t-2} B_{t-1}, C_{t-2} C_{t-1})$,
where $A_t$, $B_t$, and $C_t$ represent the moves taken from $\{c,d\}$
by Alice, Bob, and Charlie at time $t$, respectively. The number of states is
thus $2^{3m} = 64$, but the actual number can be reduced to $40$ because
Alice's moves will not be affected even if Bob and Charlie exchange
their names (Table~\ref{tab:abbreviated_notations}). For this reason, the number
of possible strategies for Alice amounts
to $N(m=2) = 2^{40} \approx 1.1 \times 10^{12}$.
This is an upper bound for direct enumeration because
the number increases to $N (m=3) = 2^{288} \approx 5.0 \times
10^{86}$, which is comparable to the estimated number of protons in the
universe. For this reason, we begin by restricting ourselves to $m=2$.

\begin{table}
  \caption{
    Abbreviated notations for states with $m=2$. The numbers on the left column
    mean how many players defected at $t-2$ and $t-1$ among Bob and Charlie,
    respectively.
  }
  \label{tab:abbreviated_notations}
  {\centering
  \begin{tabular}{cc}
    \hline
    Abbreviated & Complete notation \\
    \hline
    $(**,22)$         & $(**,dd,dd)$ \\
    $(**,21)$         & $(**,dd,dc)$ or $(**,dc,dd)$ \\
    $(**,20)$         & $(**,dc,dc)$ \\
    $(**,12)$         & $(**,cd,dd)$ or $(**,dd,cd)$ \\
    $(**,11)$         & $(**,cd,dc)$ or $(**,dc,cd)$ \\
    $(**,1\underbar{$1$})$   & $(**,dd,cc)$ or $(**,cc,dd)$ \\
    $(**,10)$         & $(**,dc,cc)$ or $(**,cc,dc)$ \\
    $(**,02)$         & $(**,cd,cd)$ \\
    $(**,01)$         & $(**,cc,cd)$ or $(**,cd,cc)$ \\
    $(**,00)$         & $(**,cc,cc)$ \\
    \hline
  \end{tabular}
  \par}
\end{table}

We are now ready to deal with the criteria. Suppose that Alice is using a
certain strategy, $s_{\text{Alice}}$. Among all the transitions between every
pair of states, only some are allowed by $s_{\text{Alice}}$:
Note that $S_{t+1} = ( A_{t-1} A_{t}, B_{t-1} B_{t}, C_{t-1} C_{t})$
shares $A_{t-1}$, $B_{t-1}$, and $C_{t-1}$ with $S_t$. From Alice's point of
view, her strategy $\Omega$ has already determined $A_t$ from $S_t$, leaving
only two
unknowns, $B_t$ and $C_t$. Therefore, every state can be followed by one of four
possibilities, depending on Bob's and Charlie's moves. In graph-theoretic terms,
each state can be mapped to a node so that every possible transition from $S_t$
to $S_{t+1}$ allowed by her strategy $s_{\text{Alice}}$ is completely specified
by a graph of $64$ nodes, each of which has $4$ outgoing links as shown in
Fig.~\ref{fig:three_conditions_sample_graph}(b).
Hereafter, we denote it as the transition graph of the strategy.
The defensibility criterion requires that Alice must not be
exploited repeatedly by any of her co-players.
Let us define a loop as a sequence of consecutive states $S_t \rightarrow
S_{t+1} \rightarrow \ldots \rightarrow S_{t+\nu}$ with $S_t = S_{t+\nu}$. This
is an important unit of analysis because only states on a loop can be visited
repeatedly to affect the players' long-term payoffs.
For $s_{\text{Alice}}$ to be defensible, it should satisfy
inequalities $\sum_{\tau=0}^{\nu-1} \left[
  F_{{\text{Alice}}}^{(\tau)}-F_{\text{Bob}}^{(\tau)} \right] \geq 0$ and
$\sum_{\tau=0}^{\nu-1} \left[
F_{\text{Alice}}^{(\tau)}-F_{\text{Charlie}}^{(\tau)} \right] \geq 0$ against
any finite-memory strategies $j$ and $k$, where the payoffs are
evaluated along every possible loop $S_t \to S_{t+1} \to \dots \to
S_{t+\nu}$ of the transition graph of $s_{\text{Alice}}$.
In other words, Alice's strategy must not contain a `risky' loop,
along which the sum of Alice's payoffs is smaller than that of either Bob or
Charlie.
If no risky loop exists in the transition graph of $s_{\text{Alice}}$,
this is a sufficient condition for $f_{\text{Alice}} \ge
f_{\text{Bob}}$ and $f_{\text{Alice}} \ge f_{\text{Charlie}}$ with arbitrary
strategies of Bob and Charlie when $e \rightarrow 0^+$~\cite{yi2017combination}.
The existence of risky loops can be investigated by means of the Floyd-Warshall
algorithm~\cite{hougardy2010floyd}.

To reduce the strategy space to search, we first check the defensibility under
the assumption that Bob or Charlie always defects (AllD). Then, we can exclude
the following states from consideration, $(**,20)$, $(**,11)$, $(**,10)$,
$(**02)$, $(**,01)$, and $(**,00)$, because these are inconsistent with the
assumption. We are left with $16$ states originating from $(**,22)$, $(**,21)$,
$(**,12)$, and $(**,1\bar{1})$. The number of possible (sub-)strategies is thus
$2^{16} = 65,536$, which is readily tractable.
By an exhaustive search for the defensibility, we obtain $48$ sub-strategies out
of the $2^{16}$ possibilities.
As a consequence, the number of strategies is reduced to $48 \times 2^{24} =
805,306,368$.
For these remaining strategies, we comprehensively check the defensibility
criterion by using a supercomputer
without the assumption that Bob or Charlie is an AllD player.

The efficiency and distinguishability criteria can be checked by
calculating $f_{\text{Alice}}$ when $\mathcal{P} = \{s_{\text{Alice}},
s_{\text{Bob}}, s_{\text{Charlie}} \} = \{s_{\text{Alice}}, s_{\text{Alice}},
s_{\text{Alice}} \}$ and $\{ s_{\text{Alice}}, \text{AllC}, \text{AllC} \}$,
respectively. The long-term payoff $f_{\text{Alice}}$ is calculated from
the stationary probability distribution over the states, which can be obtained
by linear algebraic calculation.
If $s_{\text{Alice}}$ fulfills all these criteria, it is an
$\Omega$ strategy, and we will also call it `successful'.

\section*{Result}

\subsection*{Successful strategies for the three-person game}

Our first result is impossibility: No memory-$2$ strategy satisfies the
efficiency and defensibility criteria together, according to our direct
enumeration of $N(m=2) = 1.1 \times 10^{12}$ cases.
Although $3,483,008$ strategies have passed the defensibility criterion,
none of them satisfies the efficiency criterion.
The joint application of defensibility and efficiency turns out to be
too tough for strategies with $m \leq 2$.

However, we have a class of strategies that are \emph{partially} efficient.
Out of $3,483,008$ defensible strategies, $544$ strategies show stationary
probability $\approx 1/8$ at $(cc,cc,cc)$, whereas the probability is close to
zero for the others.
We further impose the distinguishability criterion and obtained
$256$ strategies that are defensible, distinguishable, and partially efficient.
Each of them will be called a partially successful strategy (PS2),
and their full list is given in Table~\ref{tab:partially_successful}.

\begin{table*}[htbp]
  \caption{
    Partially successful strategies with $m=2$. For five different states,
    the moves are indicated by $*$, which means that they can be arbitrarily
    chosen from $\{c,d\}$.
    Some other states have multiple $A_t$'s, which may be denoted as $[A_t^{(1)}
    A_t^{(2)}\ldots A_t^{(8)}]$. One must choose moves with the same upper
    index at these states, so they give $8$ possibilities.
    The total number of strategies covered here is thus $2^5 \times 8 = 256$.
    Every third column shows which rule explains the move (see Section 4).
  }
  \label{tab:partially_successful}
  {\centering
  \begin{tabular}{|ccc|ccc|ccc|ccc|}
    \hline
    State & $A_t$ & Rule & State & $A_t$ & Rule & State & $A_t$ & Rule & State & $A_t$ & Rule \\
    \hline
    $(cc,00)$              & $c$          & i  & $(cd,00)$              & $c$          & ii & $(dc,00)$              & $d$          & ii  & $(dd,00)$              & $c$        & vi \\
    $(cc,01)$              & $d$          &    & $(cd,01)$              & $*$          &    & $(dc,01)$              & $[ddcdcdcd]$ &     & $(dd,01)$              & $*$        &    \\
    $(cc,02)$              & $d$          &    & $(cd,02)$              & $[ddccdddd]$ &    & $(dc,02)$              & $c$          & iii & $(dd,02)$              & $d$        &    \\
    $(cc,10)$              & $[cdccccdd]$ & i  & $(cd,10)$              & $d$          &    & $(dc,10)$              & $c$          & i   & $(dd,10)$              & $d$        &    \\
    $(cc,11)$              & $d$          &    & $(cd,11)$              & $c$          & iv & $(dc,11)$              & $d$          &     & $(dd,11)$              & $*$        &    \\
    $(cc,1\underbar{$1$})$ & $d$          &    & $(cd,1\underbar{$1$})$ & $d$          &    & $(dc,1\underbar{$1$})$ & $d$          &     & $(dd,1\underbar{$1$})$ & $d$        &    \\
    $(cc,12)$              & $[ccdddddd]$ &    & $(cd,12)$              & $d$          &    & $(dc,12)$              & $d$          &     & $(dd,12)$              & $d$        &    \\
    $(cc,20)$              & $c$          & i  & $(cd,20)$              & $d$          &    & $(dc,20)$              & $*$          & i   & $(dd,20)$              & $d$        &    \\
    $(cc,21)$              & $d$          &    & $(cd,21)$              & $d$          &    & $(dc,21)$              & $c$          & v   & $(dd,21)$              & $d$        &    \\
    $(cc,22)$              & $d$          &    & $(cd,22)$              & $d$          &    & $(dc,22)$              & $*$          &     & $(dd,22)$              & $d$        &    \\
    \hline
  \end{tabular}
  \par}
\end{table*}

\begin{figure*}
\begin{center}
  \includegraphics[width=1.5\columnwidth]{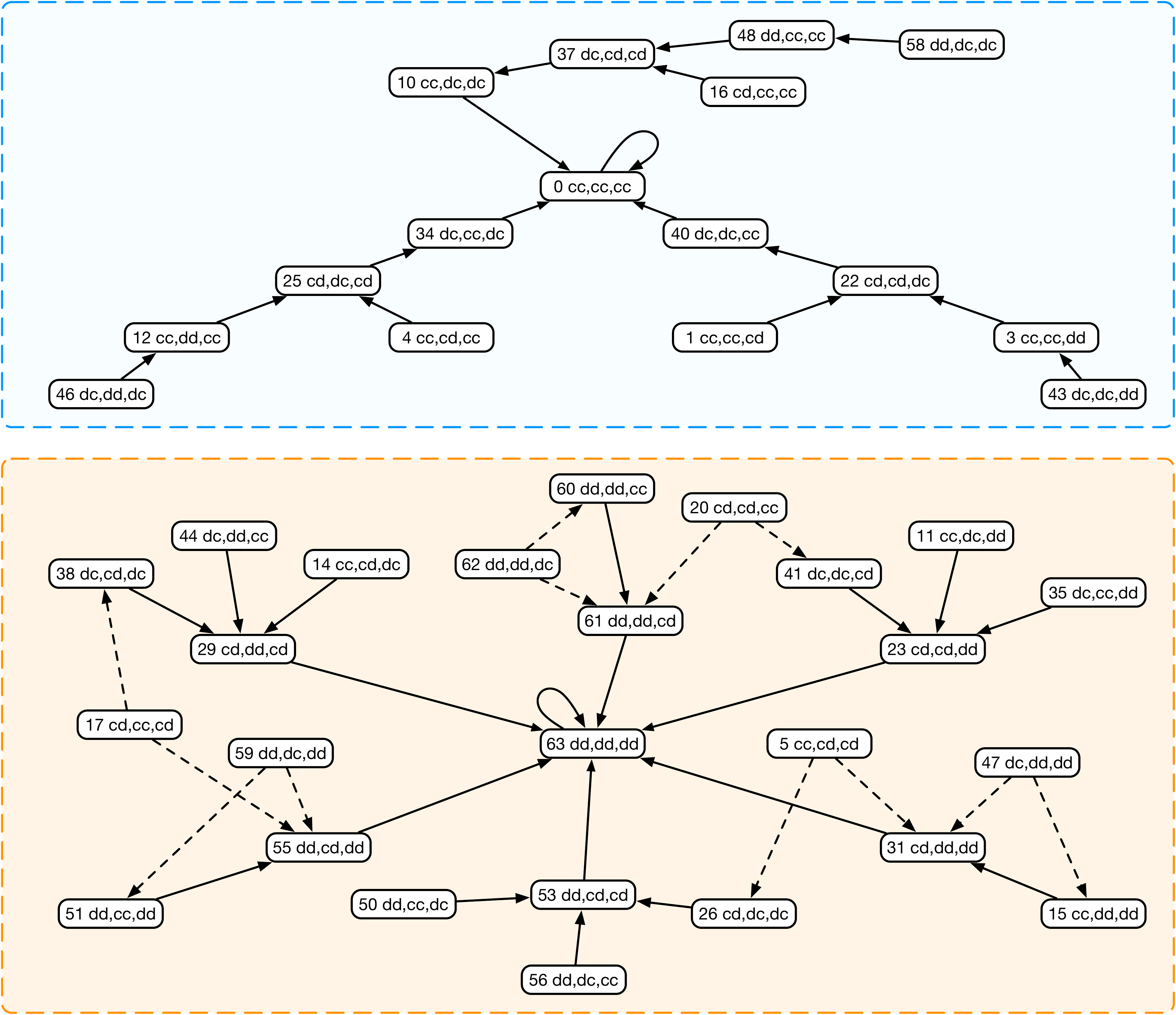}
  \caption{\label{fig:itg_partially_successful}
    Common subgraph of the $256$ partially successful strategies with $m=2$,
    when all three players adopt the same strategy.
    The label of each node indicates the state in the form of
    $A_{t-2}A_{t-1},B_{t-2}B_{t-1},C_{t-2}C_{t-1}$ prefixed by an index ranging
    from $0$ to $63$.
    The connected components including the fully cooperative state $(cc,cc,cc)$
    and the fully defective state $(dd,dd,dd)$ are depicted by light-blue and
    orange boxes, respectively.
    Some nodes have double dotted links instead of a single solid link,
    indicating that the outgoing link depends on
    which strategy to choose among the $256$ strategies.
  }
  \end{center}
\end{figure*}

The low efficiency of a PS2 is explained as follows:
When all players adopt a PS2, some states converge to $(cc,cc,cc)$ and some
others to $(dd,dd,dd)$. The respective sets of the
states will be denoted as $c$- and $d$-clusters
(Fig.~\ref{fig:itg_partially_successful}).
We note that the fully defective state $(dd,dd,dd)$ is robust against one-bit
error because $(dc,dd,dd)$, $(dd,dc,dd)$, and $(dd,dd,dc)$ belong to the
$d$-cluster. It is actually a necessary condition to be defensible
against AllD: A
player must defect when one of the co-players keeps defecting. Suppose that they
are trapped in the fully defective state. Even if Charlie cooperates by mistake,
Alice and Bob have no chance to change their moves because these two cannot
distinguish each other from an AllD player. When the same argument applies to
the $n$-person game, the fully defective state must be robust against
$(n-2)$-bit error. To escape from $(dd,dd,dd)$, two players, say Bob and
Charlie, have to
make error at the same time. Then, Alice may turn to cooperation at a subsequent
round. As a consequence, the probability to escape from the $d$-cluster, denoted
as $P_{\rm esc}^{(d)}$, is of $O(e^2)$ for $n=3$. If we look at the probability
of escaping from the $c$-cluster, $P_{\rm esc}^{(c)}$, it also turns out to be
of $O(e^2)$. Because the escape probabilities have the same order of magnitude,
the system can transit back and forth between the $c$- and $d$-clusters, so that
the clusters occupy similar amounts of stationary probabilities even in the
limit of $e \to 0$. This is the reason that the stationary probability of
$(cc,cc,cc)$ is significantly less than $100\%$.

\begin{figure*}
\begin{center}
  \includegraphics[width=1.5\columnwidth]{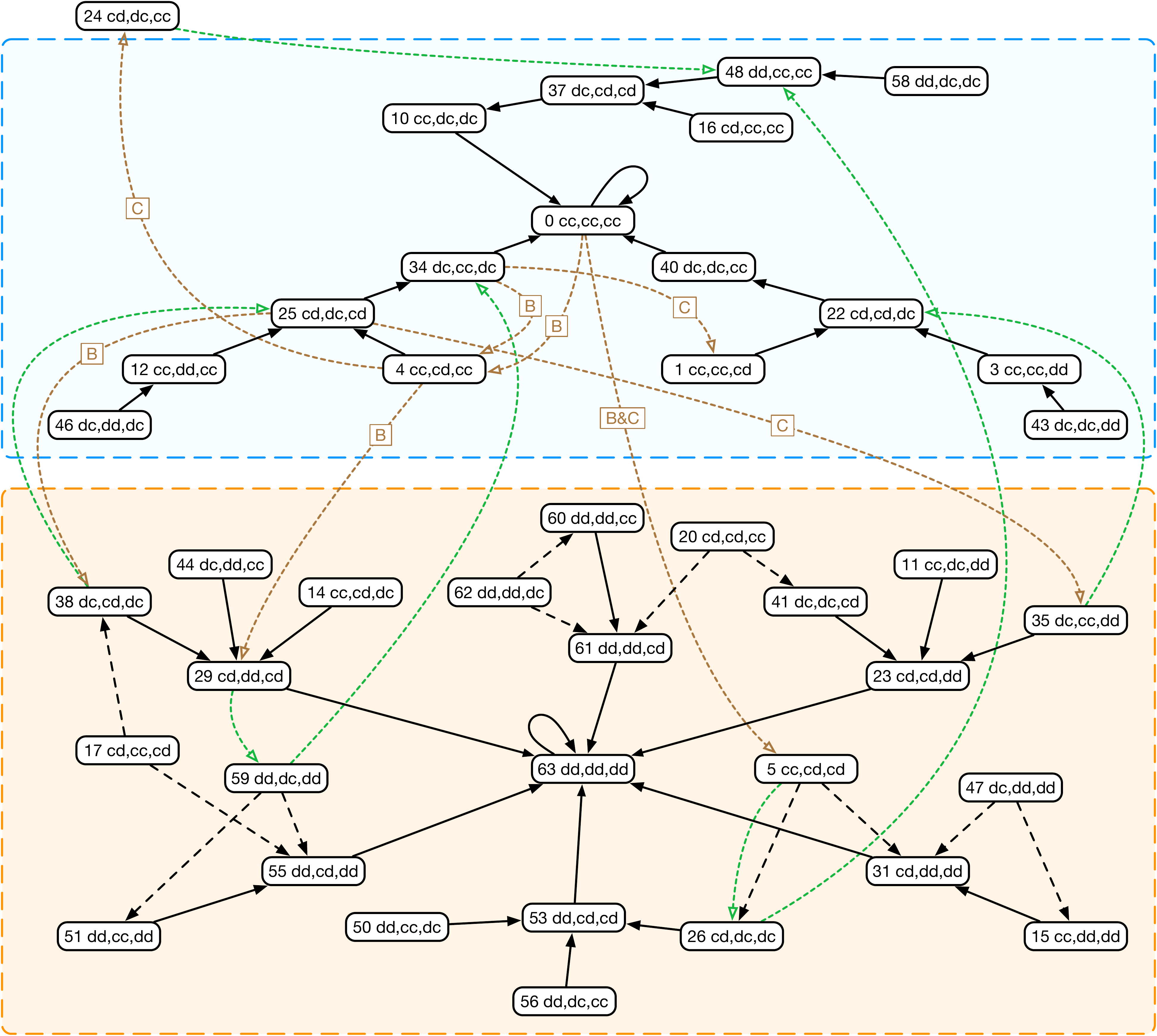}
  \caption{\label{fig:itg_partially_successful_m3}
  Modified subgraphs with $m=3$.
  The paths along which the states change by one- or two-bit error are depicted
  by brown arrows.
  Only Bob's and Charlie's implementation errors are drawn
  because we have to consider errors up to $O(e^2)$ and the graph is symmetric
  with respect to Alice, Bob and Charlie.
  Characters ``B'' and ``C'' shown on these arrows indicate whose move was
  altered by the error.
  There are five possible paths to go outside of the connected component: $(0
  \to 5)$, $(4 \to 29)$, $(4 \to 24)$, $(25 \to 38)$, and $(25 \to 35)$.
  The green paths are introduced for $m=3$
  successful strategies to recover from these noise.
  These paths are taken only when the state changes according to the brown
  arrows
  in previous rounds. For instance, the move at node $38$ is usually $d$ but
  it becomes $c$ only when the previous node was $25$. By introducing green
  arrows, the state returns to the fully cooperative node.
  }
  \end{center}
\end{figure*}

\begin{table}
  \caption{
    Moves added to partially successful strategies to make them satisfy the
    efficiency condition. The states with $m=3$ are denoted as
    $(A_{t-3}A_{t-2}A_{t-1},B_{t-3}B_{t-2}B_{t-1},C_{t-3}C_{t-2}C_{t-1})$.
    Alice must choose $c$ at these states.
  }
  \label{tab:m3_override_moves}
  {\centering
  \begin{tabular}{|cc|cc|}
    \hline
    State           & $A_t$ &
    State           & $A_t$ \\ \hline
    $(ccd,ccd,ccc)$ & $c$ & $(cdc,ccd,ccc)$ & $c$ \\
    $(ccd,ccc,ccd)$ & $c$ & $(cdc,ccc,ccd)$ & $c$ \\
    $(cdc,cdc,ccd)$ & $c$ & $(ccc,cdc,ccd)$ & $c$ \\
    $(cdc,ccd,cdc)$ & $c$ & $(ccc,ccd,cdc)$ & $c$ \\
    $(cdd,ccd,ccd)$ & $c$ & $(dcd,cdc,cdc)$ & $c$ \\
    $(ddc,cdd,cdd)$ & $c$ & $(cdd,dcc,cdc)$ & $c$ \\
    $(cdd,ddc,cdd)$ & $c$ & $(cdd,cdc,dcc)$ & $c$ \\
    $(cdd,cdd,ddc)$ & $c$ &                 &     \\
    \hline
  \end{tabular}
  \par}
\end{table}

To make the strategy efficient, the $c$-cluster must be robust against any
two-bit error, i.e., yielding $P_{\rm esc}^{(c)} \sim O(e^3)$.
Our finding is that it is possible to design successful strategies by making the
memory length longer and overriding some of the moves prescribed by the PS2.
We have enumerated all possible occurrences of two-bit errors and introduced
moves to correct these errors as shown in Table~\ref{tab:m3_override_moves}.
In Fig.~\ref{fig:itg_partially_successful_m3}, we depict paths due to two-bit
flip errors with brown arrows.
To recover mutual cooperation, we add recovery paths as indicated by the green
arrows.
The strategy goes as follows:
(i) Each player will usually follow one of the PS2's. (ii) If the memory of
three consecutive states shows unusual transition such as represented by the
brown arrows, the players will activate ``Plan B'' to follow the green arrows.
In other words, we override the moves in Table~\ref{tab:m3_override_moves},
whereby the memory-2 PS2's are extended to memory-3 successful strategies.
We have confirmed that the stationary probability of the fully cooperative state
$(ccc,ccc,ccc)$ indeed approaches one as $e \to 0$ without violating the
defensibility and distinguishability criteria.
It is thus concluded that at least $256$ successful strategies do exist for the
three-person PG game when memory length is three.
One of the successful memory-3 strategies thereby obtained is shown in Table~\ref{tab:m3_successful_strategy}.

\begin{table*}
\caption{One of successful memory-3 strategies. We have picked up the strategy
having the largest number of $c$.
  The left column shows the state of Bob and Charlie, whereas Alice's state is shown on the right.
}
\label{tab:m3_successful_strategy}
{\centering
\begin{tabular}{c|cccccccc}
  \hline
  & \multicolumn{8}{c}{$A_{t-3}A_{t-2}A_{t-1}$} \\
  $B_{t-3}B_{t-2}B_{t-1}C_{t-3}C_{t-2}C_{t-1}$ & $ccc$ & $ccd$ & $cdc$ & $cdd$ & $dcc$ & $dcd$ & $ddc$ & $ddd$ \\
  \hline
$cccccc$ & $c$ & $c$ & $d$ & $c$ & $c$ & $c$ & $d$ & $c$ \\
$cccccd$ / $ccdccc$ & $d$ & $c$ & $c$ & $c$ & $d$ & $c$ & $c$ & $c$ \\
$ccccdc$ / $cdcccc$ & $c$ & $d$ & $c$ & $d$ & $c$ & $d$ & $c$ & $d$ \\
$ccccdd$ / $cddccc$ & $d$ & $d$ & $d$ & $d$ & $d$ & $d$ & $d$ & $d$ \\
$cccdcc$ / $dccccc$ & $c$ & $c$ & $d$ & $c$ & $c$ & $c$ & $d$ & $c$ \\
$cccdcd$ / $dcdccc$ & $d$ & $c$ & $c$ & $c$ & $d$ & $c$ & $c$ & $c$ \\
$cccddc$ / $ddcccc$ & $c$ & $d$ & $c$ & $d$ & $c$ & $d$ & $c$ & $d$ \\
$cccddd$ / $dddccc$ & $d$ & $d$ & $d$ & $d$ & $d$ & $d$ & $d$ & $d$ \\
$ccdccd$            & $d$ & $c$ & $c$ & $c$ & $d$ & $c$ & $c$ & $d$ \\
$ccdcdc$ / $cdcccd$ & $c$ & $c$ & $c$ & $c$ & $d$ & $c$ & $d$ & $c$ \\
$ccdcdd$ / $cddccd$ & $d$ & $d$ & $d$ & $d$ & $d$ & $d$ & $d$ & $d$ \\
$ccddcc$ / $dccccd$ & $d$ & $c$ & $c$ & $c$ & $d$ & $c$ & $c$ & $c$ \\
$ccddcd$ / $dcdccd$ & $d$ & $c$ & $c$ & $d$ & $d$ & $c$ & $c$ & $d$ \\
$ccdddc$ / $ddcccd$ & $d$ & $c$ & $d$ & $c$ & $d$ & $c$ & $d$ & $c$ \\
$ccdddd$ / $dddccd$ & $d$ & $d$ & $d$ & $d$ & $d$ & $d$ & $d$ & $d$ \\
$cdccdc$            & $c$ & $d$ & $c$ & $d$ & $c$ & $c$ & $c$ & $d$ \\
$cdccdd$ / $cddcdc$ & $d$ & $d$ & $c$ & $d$ & $d$ & $d$ & $c$ & $d$ \\
$cdcdcc$ / $dcccdc$ & $c$ & $d$ & $c$ & $c$ & $c$ & $d$ & $c$ & $d$ \\
$cdcdcd$ / $dcdcdc$ & $d$ & $c$ & $d$ & $c$ & $d$ & $c$ & $d$ & $c$ \\
$cdcddc$ / $ddccdc$ & $c$ & $d$ & $c$ & $d$ & $c$ & $d$ & $c$ & $d$ \\
$cdcddd$ / $dddcdc$ & $d$ & $d$ & $c$ & $d$ & $d$ & $d$ & $c$ & $d$ \\
$cddcdd$            & $d$ & $d$ & $c$ & $d$ & $d$ & $d$ & $c$ & $d$ \\
$cdddcc$ / $dcccdd$ & $d$ & $d$ & $d$ & $d$ & $d$ & $d$ & $d$ & $d$ \\
$cdddcd$ / $dcdcdd$ & $d$ & $d$ & $d$ & $d$ & $d$ & $d$ & $d$ & $d$ \\
$cddddc$ / $ddccdd$ & $d$ & $d$ & $c$ & $c$ & $d$ & $d$ & $c$ & $d$ \\
$cddddd$ / $dddcdd$ & $d$ & $d$ & $c$ & $d$ & $d$ & $d$ & $c$ & $d$ \\
$dccdcc$            & $c$ & $c$ & $d$ & $c$ & $c$ & $c$ & $d$ & $c$ \\
$dccdcd$ / $dcddcc$ & $d$ & $c$ & $c$ & $c$ & $d$ & $c$ & $c$ & $c$ \\
$dccddc$ / $ddcdcc$ & $c$ & $d$ & $c$ & $d$ & $c$ & $d$ & $c$ & $d$ \\
$dccddd$ / $ddddcc$ & $d$ & $d$ & $d$ & $d$ & $d$ & $d$ & $d$ & $d$ \\
$dcddcd$            & $d$ & $c$ & $c$ & $d$ & $d$ & $c$ & $c$ & $d$ \\
$dcdddc$ / $ddcdcd$ & $d$ & $c$ & $d$ & $c$ & $d$ & $c$ & $d$ & $c$ \\
$dcdddd$ / $ddddcd$ & $d$ & $d$ & $d$ & $d$ & $d$ & $d$ & $d$ & $d$ \\
$ddcddc$            & $c$ & $d$ & $c$ & $d$ & $c$ & $d$ & $c$ & $d$ \\
$ddcddd$ / $dddddc$ & $d$ & $d$ & $c$ & $d$ & $d$ & $d$ & $c$ & $d$ \\
$dddddd$            & $d$ & $d$ & $c$ & $d$ & $d$ & $d$ & $c$ & $d$ \\
\hline
\end{tabular}
\par}
\end{table*}

\subsection*{Necessary memory length for the $n$-person game}

Generalizing the above impossibility result, we can show that $m \geq
n$ is required for a strategy to be successful for the $n$-players PG game when $n \geq 3$.
We have already seen that the fully defective state must be
robust against $(n-2)$-bit error to be defensible against AllD, which means
that $P_{\rm esc}^{(d)} \lesssim O(e^{n-1})$.
On the other hand, the efficiency criterion requires that
$P_{\rm esc}^{(c)} / P_{\rm esc}^{(d)} \to 0$ as $e \to 0$.
In other words, we need $P_{\rm esc}^{(c)} \lesssim O(e^{n})$, which
implies that the fully cooperative state has to be robust against $(n-1)$-bit
error. We note the following: If the fully
cooperative state of a memory-$m$ strategy is robust against $k$-bit error,
its memory length $m$ must be greater than $k$ for this strategy to be
defensible. A rigorous proof for this statement is given in the next paragraph,
but a rough explanation goes as follows: Suppose that $k$-bit error happened to
a player, say Bob, so that he took the opposite moves $k$ times in a row by
mistake. He must have $m=k+1$ at least to realize and correct his own mistakes.
Otherwise, he could not tell if he has committed the errors. In our case, $k$
equals $n-1$ in the $n$-person PG game, which leads to the inequality $m \ge n$.
In what follows, we will show that a memory-$k$ strategy cannot satisfy
the defensibility criterion if it makes the fully cooperative state robust
against $k$-bit error. The proof consists of two steps:

\begin{figure}
  \caption{
    (a) Basic notations.
    We assume that strategy $S$ recovers cooperation from $k$ successive
    errors when all three players have adopted it.
    All the players were cooperating at $t < 0$.
    For $1 \leq t \leq k$, Alice has at most $k$ sequential errors.
    By assumption, $S$ is robust against $k$-bit error, so all the players
    must recover full cooperation at a certain time step, $t_{\rm rec}$.
    The sequences of Alice and the others are denoted as $\Gamma$ and $\Delta$,
    respectively.
    (b) Breaking the defensibility of $S$. The group of $A$'s co-players,
    denoted as
    $\bar{A}$, start defecting and then simulate a proper move sequence to
    recover full cooperation.
  }
  \label{fig:error_notation}
  \begin{center}
  \includegraphics[width=0.9\columnwidth]{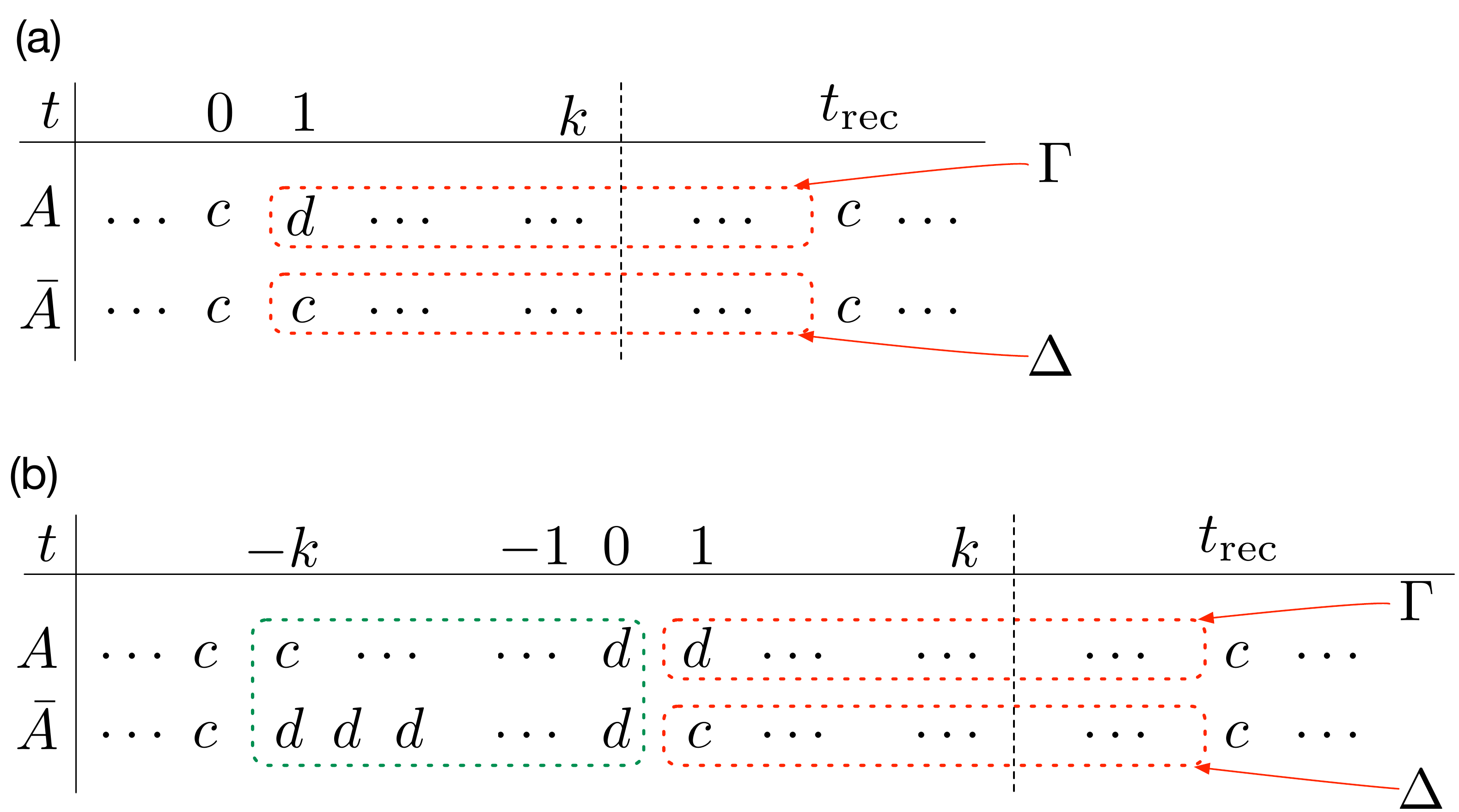}
  \end{center}
\end{figure}

First,
suppose that there exists a memory-$k$ defensible strategy $S$ whose fully
cooperative state is robust against $k$-bit error.
We begin by assuming that the players are in the fully cooperative
state with a strategy profile $\mathcal{P} = \{S, S, \ldots, S\}$.
If error occurs at $t=1$ and may also have occurred for $2 \leq t \leq
k$ only on the moves of Alice (denoted by $A$), the fully cooperative
state is recovered in a finite time step $t=t_{\rm rec} ( > k)$.
This is because the fully cooperative state of the strategy $S$ is assumed to be
robust against $k$-bit error and the total number of errors is less than or
equal to $k$.
Depending on how error occurs for $2 \leq t \leq k$, the sequences of moves
taken by $A$ for this period can be arbitrary, so the number of possible
patterns is $2^{k-1}$. Each of these sequence of Alice's moves will be
denoted by $\Gamma_{i}$, where $i = 1 \dots 2^{k-1}$. The move at $t$ in
$\Gamma_{i}$ is denoted by $\Gamma_{i}^{t}$, which is either $c$ or $d$.
On the other hand, we suppose that
no error occurs on the other players (denoted by $\bar{A}$ as a collective
entity), so $\bar{A}$ shows an identical sequence of moves, following $S$.
The sequence of moves by $\bar{A}$ for the noise pattern $i$ is denoted by
$\Delta_i$ and its move at $t$ is denoted by $\Delta_i^t$
(Fig.~\ref{fig:error_notation}(a)).
It is important that
$\Delta_i^t$ actually depends only on $A$'s previous moves $\Gamma_i^1, \dots,
\Gamma_i^{t-1}$ because the moves of $\bar{A}$ are deterministic.
Let $P( \Gamma_i )$ and $P( \Delta_i )$ denote the payoff of $A$ and $\bar{A}$
for $1 \leq t \leq t_{\rm rec}$, respectively.
According to the defensibility of $\bar{A}$'s strategy $S$,
we have an inequality $P( \Gamma_i ) \leq P( \Delta_i )$, whereas
$P( \Gamma_i ) \geq P( \Delta_i )$ is not necessarily true
because $A$ made errors in following the strategy $S$.

Second, we consider the case where $A$ follows the strategy $S$
but the other players in $\bar{A}$ make an alliance and move together
(Fig.~\ref{fig:error_notation}(b)). We will
show that $\bar{A}$ can repeatedly exploit $A$ by choosing their moves as
follows.
\begin{enumerate}
  \item Start from full cooperation.
  \item $\bar{A}$ defects first, while $A$ is cooperating.
  \item $\bar{A}$ continues defecting until they reach full defection.
    $A$ must eventually defect, otherwise she is exploited.
    At this stage, $\bar{A}$ has a higher net payoff than $A$'s because they
    started defection earlier than $A$.
  \item At a certain point, say $t=1$, $\bar{A}$ returns to $c$.
    $A$ is still defecting to defend herself, so we observe $(\Gamma_i^1,
    \Delta_i^1) = (d,c)$.
  \item $\bar{A}$ then takes $\Delta_i^2$ of Fig.~\ref{fig:error_notation}(a).
    Note that $\Delta_i^2$ is independent of $i$ because it only depends on
    $\Gamma_i^1$ and $\Delta_i^1$, which are fixed as $d$ and $c$, respectively.
    On the other hand, $A$'s move, $\Gamma_i^2$, may be either $c$ or $d$,
    depending on $S$.
  \item Find a sequence $i$ for which $\Gamma_i^2$ equals $A$'s previous move.
    $\bar{A}$ then takes $\Delta_i^3$, which is identical for any $\Delta_i$
    as long as $\Gamma_i^2$ is the same.
    Once again, $A$'s move, $\Gamma_i^3$, may be either $c$ or $d$.
  \item Repeat the above sequence until $\bar{A}$ takes
  $\Delta_i^{t_{\rm rec}}$. In short, $\bar{A}$ is simulating one of
  $\Delta_i$'s to recover full cooperation.
  Which $\Delta_i$ to choose depends on $S$, but there is
  always one noise pattern that produces such $\Gamma_i$ and $\Delta_i$.
  After this series of moves, they eventually get back to full cooperation.
\end{enumerate}
In addition to the payoff advantage in step 3,
$\bar{A}$'s net payoff from step 4 to 7 is always higher than or equal to
$A$'s because $P( \Delta_i ) \geq P( \Gamma_i )$ for any $i$.
In this way, $\bar{A}$ can repeatedly exploit $A$, which
contradicts our assumption that $S$ is defensible. Hence, there is no
memory-$k$ strategy $S$ that is defensible and robust against $k$-bit error.
For this reason, if the fully cooperative state of
a memory-$m$ strategy is robust against $k$-bit error, its memory length $m$
must be greater than $k$ for this strategy to be defensible.
If Alice's memory length is longer than $k$, on the other hand,
the state from Alice's point
of view will be $(d, \Gamma_i^1, \Gamma_i^2,  \ldots,
\Gamma_i^k, d, \Delta_i^1, \Delta_i^2, \ldots, \Delta_i^k)$.
She can thus recognize that the state started from full defection and
that Bob and Charlie initiated the change. In this case, Alice can defend
herself by keeping $d$.

\section*{Discussion}

In summary, we have found that three players can safely maintain full
cooperation in the PG game in a noisy environment with $e \to 0^+$.
From Tables~\ref{tab:partially_successful} and \ref{tab:m3_override_moves}, we
see that a successful strategy $\Omega$ is conditioned by the following rules.
\begin{enumerate}
\item Preserve cooperation. If everyone cooperated last time, Alice
chooses $c$ according to $\Omega$.
In other words, $c$ is the choice at $(*c,*0)$, such as
$(cc,00)$, $(cc,10)$, $(cc,20)$, $(dc,10)$, and $(dc,20)$
(see Table~\ref{tab:abbreviated_notations} for the abbreviated notations).
\item Challenge the co-players' naivety. An exception of the first rule is
$(dc,00)$, at which $d$ is prescribed because of the distinguishability
criterion. Due to this prescription, if Bob and Charlie are AllC, they are
exploited by Alice who alternates between $(dc,00)$ and $(cd,00)$. This explains
why Alice cooperates at $(cd,00)$.
\item Retreat if the co-players are not naive.
If Bob and Charlie are not AllC, on the other hand, they will choose $d$
in response to Alice's unilateral defection, so the resulting state will be
$(dc,02)$ instead of $(dc,00)$. If this is the case, Alice cooperates to avoid
TFT retaliation.
\item Forgive after responding to provocation.
Note that $(dc,02)$ corresponds to $(dc,cd,cd)$, which Bob or
Charlie would interpret as $(cd,11)$. The strategy prescribes $c$ at $(cd,11)$
so that full cooperation is recovered quickly when Bob and Charlie also use
$\Omega$.
\item Grab the chance to cooperate.
If the state was initially full defection, it is definitely safe to defect, so
$(dc,21)$ is accessed by two simultaneous mistakes of Alice and another player
with probability of $O(e^2)$. As a result, $(dc,dc,dd)$ and its
permutations form the outermost periphery of the $c$-cluster in
Fig.~\ref{fig:itg_partially_successful}. If Alice reaches $(dc,21)$ by chance,
she chooses $c$ once again to establish full cooperation.
\item Don't be evil. If everyone uses $\Omega$, $(dc,21)$ such as
$(dc,dc,dd)$ is followed by $(cc,cc,dd)$, which is $(dd,00)$ from Charlie's
viewpoint. He has been the only defector for two rounds, but he is now supposed
to contribute to the public pool. If the co-players are using $\Omega$, they
will punish Charlie's successive defection, which leads the state to
$(dc,02)$ that we have already examined at the third rule.
\end{enumerate}
The above six rules explain the basic behavior of a PS2
(Table~\ref{tab:partially_successful}). To make it fully efficient, we have to
add one more rule:
\begin{enumerate}
\setcounter{enumi}{6}
\item Look at the context.
With referring to the memory of $t-3$, one must override
some moves of a PS2 as listed in Table~\ref{tab:m3_override_moves}.
The basic recipe is that Alice has to cooperate in the
subsequent two rounds if she defected by mistake. In addition, the
transition from state $(dd,dc,dd)$ to $(dc,cc,dc)$ must also be allowed if
the former one is a part of $(cdd,ddc,cdd)$.
\end{enumerate}

We believe that most of these patterns can be extended to $n > 3$ in principle,
although we have not completed the search for a successful strategy in these
cases yet. Obviously, the necessary condition of $m \geq n$ means that the
strategy space expands super-exponentially: The number of strategies for the
$n$-person game with memory length $n$ is $2^{2^{n^2}}$, which is far beyond our
computational feasibility if we are to enumerate these strategies
comprehensively. A possible alternative could be to build a strategy based on
a successful strategy of the $(n-1)$-person game.
We are currently working on the four-person game based on a successful strategy
for $n=3$. Since a successful strategy for $n=3$ has the $c$-cluster robust
against two-bit error,
we will be able to construct a PS2 for $n=4$ based on them, whose $c$- and $d$-clusters are two-bit error tolerant.
We expect that a PS2 can be elevated to successful strategies by extending the
memory length and overriding some of the moves just as we did for $n=3$.
If this attempt succeeds, it will then be possible to apply this procedure
iteratively to solve the general $n$-person game.

Similarly to the ZD strategies, our successful strategies are capable to give
the player control over the payoff differences relative to the co-players'. A
nontrivial aspect of our finding is that one may publicly announce his or her
deterministic strategy, making the future moves predictable by the co-players.
By analyzing the announced strategy, the co-players understand that it is
impossible to exploit the focal player. With the knowledge of the focal player's
strategy, the co-players may also safely adopt the same one to enjoy the above
properties, by which a cooperative Nash equilibrium is reached.
In an actual human population, of course,
it would be fair to say that the performance generally
depends on the learning dynamics as well as the
interaction structure~\cite{szolnoki2012conditional,battiston2017determinants},
and the precise understanding of their interplay will pose an interesting
research question. From a biological point of view,
one could also ask if our computational solution is accessible by means of a
certain form of
evolutionary dynamics. We believe that it is possible in principle, but it could
take an exceedingly long time because the number of strategic possibilities is
literally astronomical. As the number of players increases, most of the
prescriptions of successful strategies will appear cryptic unless one looks at
the whole connection structure of states, just as $c$ sometimes turns out to be
the correct choice for Alice at $(dc,22)$ (Table~\ref{tab:m3_override_moves}).
It suggests that our moral instinct, which has been shaped by evolution, might
fail to guide us in dealing with the immense complexity of the game. The
important point is that we can nevertheless induce mutual cooperation by
advising the players to adopt this strategy, in the light of its efficiency,
defensibility, and distinguishability.





\section*{Acknowledgment}
We gratefully acknowledge discussions with Beom Jun Kim and Hyeong-Chai Jeong.
Y.M. acknowledges support from CREST, JST and from MEXT as ``Exploratory
Challenges on Post-K computer (Studies of multi-level spatiotemporal simulation
of socioeconomic phenomena)'' and from Japan Society for the Promotion of
Science (JSPS) (JSPS KAKENHI; grant no. 18H03621).
S.K.B. was supported by Basic Science Research
Program through the National Research Foundation of Korea (NRF) funded by the
Ministry of Science, ICT and Future Planning (NRF grant no.
2017R1A1A1A05001482). This project was supported by JSPS and NRF under the
Japan-Korea Scientific Cooperation Program. This work was supported under the
framework of international cooperation program managed by the National Research
Foundation of Korea (NRF grant no. 2016K2A9A2A08003695). This research used
computational resources of the K computer provided by the RIKEN Center
for Computational Science through the HPCI System Research project
(Project ID:hp160264).

\bibliographystyle{model1-num-names}
\bibliography{pub}







\end{document}